# Optomechanical Resonating Probe for Very High Speed Sensing of Atomic Forces

Pierre Etienne ALLAIN[1], Lucien SCHWAB[2], Colin MISNER[3], Marc GELY[5], Estelle MAIRIAUX[4], Maxime HERMOUET[5], Benjamin WALTER[4], Giuseppe LEO[1], Sébastien HENTZ[5], Marc FAUCHER[3,4], Guillaume JOURDAN[5], Bernard LEGRAND[2], Ivan FAVERO[1]

1, Matériaux et Phénomènes Quantiques, Université Paris Diderot, CNRS UMR 7162, Paris, France

2, Laboratoire d'Analyse et d'Architecture des Systèmes (LAAS), CNRS UPR 8001, Université de Toulouse, Toulouse, France

3, Institut d'Electronique, de Microélectronique et de Nanotechnologie, Université de Lille, CNRS UMR 8520 - IEMN, Lille, France

4, Vmicro SAS, Avenue Poincaré, Villeneuve d'Ascq, France

5, Université Grenoble Alpes, CEA, LETI, Minatec Campus, Grenoble, France

## Abstract

Atomic force spectroscopy and microscopy (AFM) are invaluable tools to characterize nanostructures and biological systems. Most experiments, including state-of-the-art images of molecular bonds, are achieved by driving probes at their mechanical resonance. This resonance reaches the MHz for the fastest AFM micro-cantilevers, with typical motion amplitude of a few nanometres. Next-generation investigations of molecular scale dynamics, including faster force imaging and higher-resolution spectroscopy of dissipative interactions, require more bandwidth and vibration amplitudes below interatomic distance, for non-perturbative short-range tip-matter interactions. Probe frequency is a key parameter to improve bandwidth while reducing Brownian motion, allowing large signal-to-noise for exquisite resolution. Optomechanical resonators reach motion detection at $10^{-18}$ m.Hz$^{-1/2}$, while coupling light to bulk vibration modes whose frequencies largely surpass those of cantilevers. Here we introduce an optically operated resonating optomechanical atomic force probe of frequency 2 decades above the fastest functional AFM cantilevers while Brownian motion is 4 orders below. Based on a Silicon-On-Insulator technology, the probe demonstrates high-speed sensing of contact and non-contact interactions with sub-picometre driven motion, breaking open current locks for faster and finer atomic force spectroscopy.

The possibility to measure and image physical forces at the nanoscale has been a turning point in the evolution of nanoscience and nanotechnology. Since the invention of the Atomic Force Microscope (AFM), atomic force spectroscopy and microscopy have gained in functionality, precision and speed, impacting at large material science, soft and condensed matter physics, chemistry and biology[1-10]. In recently developed high-speed AFMs (HS-AFMs), the cantilever's dimensions are reduced until the detection optical spot reaches the diffraction limit, leading in practice to flexural frequencies in the MHz range and video-rate force imaging[4,8,11]. In order to ensure large signal to noise ratio in force detection, these probes are usually resonantly driven well above their Brownian motion, typically at a nanometer of amplitude[1,2,4,8]. Such amplitude, far beyond interatomic distances, hinders the optimal non-perturbative investigation of short-range interactions involved in chemical bonds[3,9,10]. Atomic force probes operating at much higher frequency and smaller amplitude, beyond these current limitations, would open a new avenue for high-speed imaging and force spectroscopy at the single molecule level.

Both speed and amplitude are impacted by a same chief parameter: the frequency $f$ of the mechanical probe[11]. The force signal is obtained by tracking $f$, which can be done within a bandwidth between $f/Q$ (with Q the probe's quality factor) and $f$ [12]. At the same time $f$ directly connects to the amplitude of Brownian motion, which gets smaller as the probe stiffens. Very high frequency mechanical probes (100 MHz and above), smaller and stiffer than existing cantilevers, hence appear as a solution. They are unfortunately difficult to control with established techniques, like electric transduction or optical spot detection.

Here, we show that a convergence of photonics, optomechanics and MEMS/NEMS technologies, based on a silicon platform, solves the above problem. We demonstrate a very high frequency ($f$>100MHz) ring-shaped atomic force probe that senses local forces exerted on its nanoscale apex. Actuated and detected fully optically thanks to strong optomechanical coupling, the probe uses a highly stiff and very high frequency bulk mechanical mode, reaching a frequency 2 decades above the fastest AFM micro-cantilevers, while motion amplitude is 4 orders below. We demonstrate the operability of the probe by sensing both contact and non-contact forces in the resonantly driven mode, showing that dissipative like non-dissipative interactions can be measured at ultra-low amplitude of motion and at very-high speed. Our results additionally indicate that the approach would be expandable to the GHz range.

## Results

The working principle of the very high-speed optomechanical atomic force probe is illustrated in Fig. 1a. The probe resonator vibrates along a very high frequency and stiff extensional (bulk) mechanical mode, which sets an apex into a quasi-1D oscillation. The vibration is actuated by an optical field stored in a laser-driven high-Q cavity, which is optomechanically coupled to the probe. The apex-to-surface elastic and dissipative interactions modulate the mechanical vibration, which is in turn transduced to the optical cavity mode. This all-optical operation of the probe, both in actuation and detection, provides ultra-fast and sensitive motion readout together with an advantageous noise management compared to electromechanical transduction[13,14]. In what follows, these assets translate into the capability to drive a very high-frequency (>100 MHz) probe with sub-picometre motion amplitude, while obtaining a detection merely limited by thermomechanical noise. To ensure compliance with instrumental applications, test chips are fabricated on a Silicon-On-Insulator (SOI) platform using a very large scale integration (VLSI) process inherited from silicon photonics (Fig. 1b). Input and output light connect through grating-couplers to an on-chip optical waveguide evanescently coupled to the optomechanical probe (Fig. 1c-d)[15,16]. In this work, we opted for a suspended silicon ring geometry, which enables intense optomechanical coupling between light travelling within the ring and the extensional modes of the structure[17,18]. The coupling strength is crucial to the all-optical mechanically-resonating operation in the very high frequency range. The nano-machined probe apex is diametrically opposed to the evanescent coupling region. The apex protrudes away from the silicon ring (Fig. 1c, 1e), allowing near-field interactions with surfaces and objects under test.

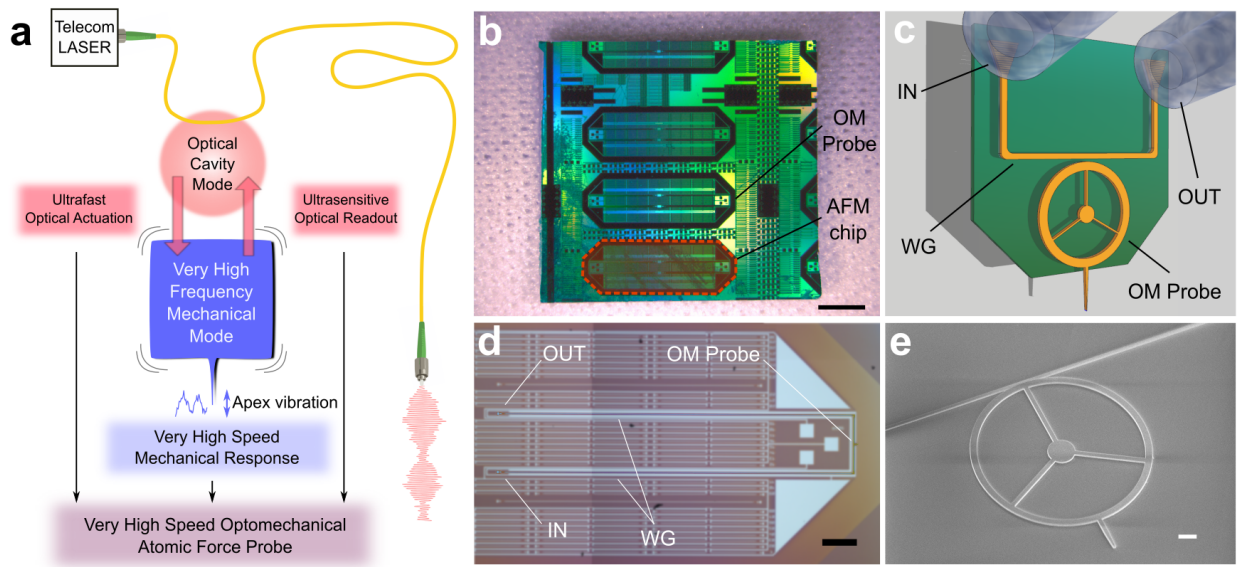

**Figure 1 | Introduction of the very high speed optomechanical atomic force probe. a,** A high-Q optical cavity mode resonating in the telecom range is coupled to a very high frequency extensional mechanical mode. The latter is optically actuated and read-out, and very rapidly responds to forces exerted onto the apex. **b,** VLSI-fabricated boat-shaped probe devices positioned in parallel on a chip, and detachable by back-etching. All devices are identical and embed one probe and its optical operation circuitry. Scale bar: 2 mm. **c,** Simplified 3D rendering of the structure highlighting the protruding apex of the optomechanical (OM) probe, the waveguide (WG) and the fibre grating couplers (IN and OUT). **d,** Zoomed optical image of a fabricated boat-shaped device. Scale bar: 300 µm. **e,** SEM micrograph of the optomechanical probe characterized in this letter. Scale bar: 4 µm.

The detailed design of the optomechanical ring is shown in Fig. S1 of the Supplementary Information. We choose a large enough ring radius (10 µm) to facilitate back etching of the substrate and allow the final probe apex protruding from the device, while we choose a ring thickness compatible with photonics standards (220 nm). The selected ring structure supports localized optical and mechanical modes (Fig. 2), which optomechanically couple via radiation pressure at a rate $g_0$~72 kHz[19], as well as via photothermal effects[20,21]. The high-Q optical mode deviates from a traditional whispering gallery mode by its non-invariant azimuthal structure (Fig. 2a). This serpentine optical mode allows positioning the bases of the apex and spokes in areas where the electromagnetic energy is minimized, thus lessening scattering optical losses (Fig. 2a). The optical mode is designed to resonate in the telecom C-band, with a wavelength around 1550 nm. Finite element simulations predict a radiative optical Q of a few $10^5$, while experiments reveal an upper limit of $Q_0$~$10^5$ with our current technology. For narrow enough apex and spokes, the optical losses are likely due to residual surface imperfections, while for wider ones ($w_{apex}$>500 nm, $w_{spk}$>500 nm), geometrical scattering losses contribute to a degradation of $Q_0$ below $10^5$. The mode employed hereafter has a resonating wavelength $\lambda_0$=1552 nm and a loaded quality factor $Q_{loaded}$~70000.

The mechanical mode is designed to maximize optomechanical coupling, as well as to reach very high frequency and thermomechanical motion well below the picometre. The radial breathing mode of the ring qualifies for these three criteria. Like the optical mode, it possesses an azimuthal modulation as a consequence of the anchoring by the spokes (Fig. 2b). Its measured frequency is $f$~117 MHz, consistent with numerical simulations, while the quality factor amounts to $Q$~2790 in vacuum. To explore the versatility of our probe design, smaller probes with a mechanical frequency of 262 MHz, as well as probes of smaller stiffness, have been characterized and are discussed in Fig. S2 of the Supplementary Information.

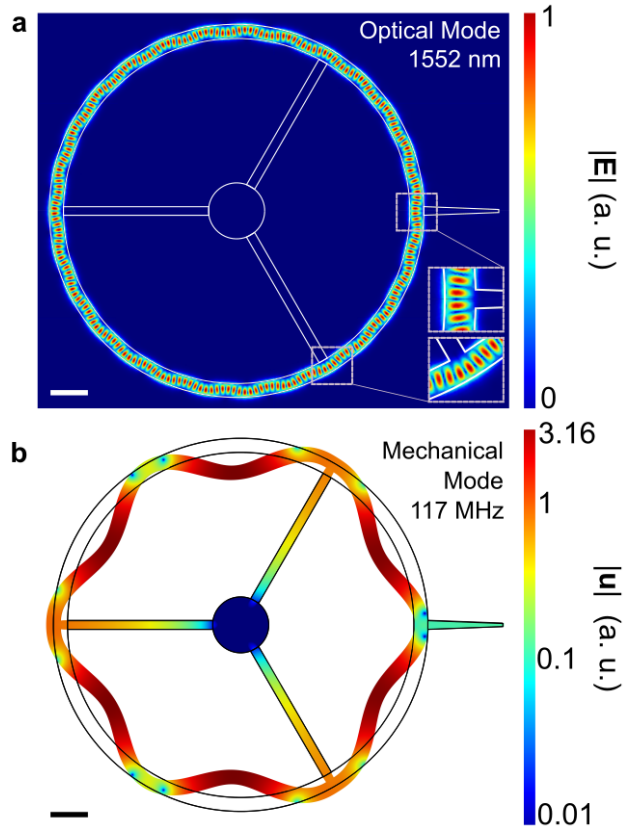

**Figure 2 | Optical and mechanical modes of the optomechanical probe. a,** Map of the normalized modulus of the electric field showing the spatial distribution of the optical cavity mode ($\lambda_0$=1552 nm, $Q_0$~$10^5$). Insets: spokes and probe apex are anchored to the ring at low field density areas, thereby minimizing scattering optical losses. **b,** Modulus of the displacement of the mechanical mode employed in this study ($f$ =117 MHz). The probe apex is displaced along its longitudinal axis, as the result of the mode's symmetry. Scale bars: 2 µm

To optically actuate the probe vibration, the light injected into the resonator is amplitude-modulated at a frequency $f_d$ close to $f$ by means of an electro-optical modulator (EOM) driven with an RF-voltage of amplitude $V_d$. The light exiting the resonator, which conveys the mechanical motion, is detected by a fast photodetector and amplified towards a lock-in amplifier (LIA). The corresponding block-diagram is detailed in Fig. S3 of the Supplementary Information. The magnitude $R_1$ of the signal vector, which is proportional to the mechanical motion amplitude, is recorded while sweeping $f_d$ around $f$ for several values of $V_d$ (Fig. 3a). The resonant vibration amplitude of the probe apex ranges from 0.1 fm to near 1 pm for the selected mechanical mode at 117 MHz, and scales linearly with the driving voltage $V_d$ (Fig. 3b). Those small amplitudes are advantageous for non-perturbative probe-matter interactions, and result from the large effective spring constant at the selected apex location. A calibration factor of 2.5 mV/pm converting the voltage output by the LIA to the probe motion amplitude was deduced by fitting the Brownian motion noise power spectrum of the mode with known parameters: temperature $T$=300 K and effective spring constant $k$~$2.6 \times 10^6$ N/m. The measured dynamic range between the Brownian and the maximally driven motion amounts to 75 dB. It is currently limited by our LIA maximal input voltage and could be further improved.

In a driven operation mode near resonance $f_d \sim f$ and under vacuum ($10^{-5}$ mbar), we characterize the performances of the probe by monitoring during 200 s the normalised shift of mechanical frequency $\Delta f/f = (f-f_0)/f_0$, with $f_0$ the unperturbed mechanical frequency (see Methods). The Allan deviation of $\Delta f/f$ is showed in Fig. 3c. For 10 s integration time, a relative deviation $\Delta f/f \sim 10^{-7}$ is routinely reached. After 10 s, experimental drifts increase the frequency noise.

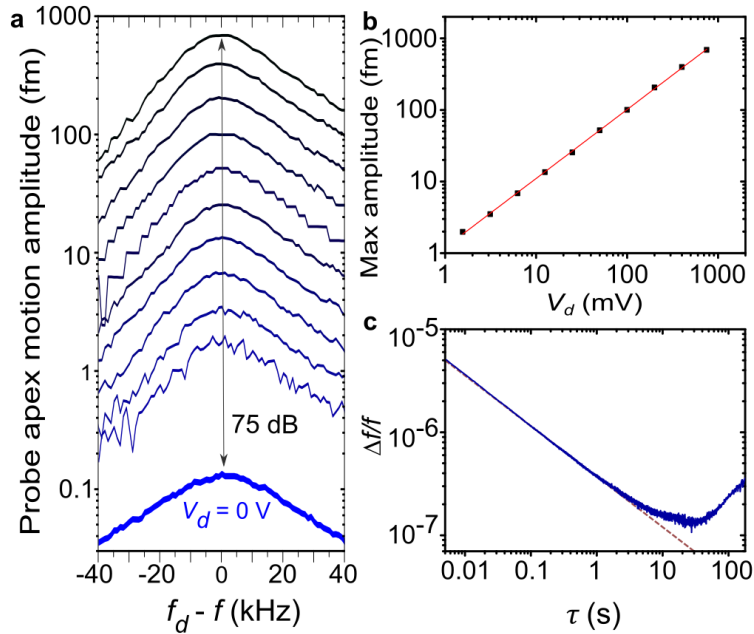

**Figure 3 | Optomechanical probe characterization. a,** Optomechanical probe apex vibration amplitude measured from the Brownian motion regime (plotted with a thicker line, BW=1 Hz) to driven amplitudes with an EOM driving voltage $V_d$ evolving between 1.5 and 750 mV. **b,** Resonant amplitude of the vibration of the apex versus drive voltage $V_d$ showing linearity along the dynamical range. **c,** Normalized Allan deviation of the mechanical frequency $\Delta f/f$ as a function of the analysis time $\tau$ (open loop).

Next we perform a benchmark approach-retract cycle (Fig. 4a), where the probe apex is approached to the tip of a separate µm-sized commercial cantilever while the probe frequency variations $\Delta f=(f-f_0)$ are monitored in a phase locked loop configuration (close-loop mode, see Supplementary Information and Fig. S3). The resulting force curve is reported in Fig. 4b. A typical hysteresis[22] cycle is evidenced, showing that the probe's apex snaps out of the cantilever tip as it is retracted. The mechanical mode frequency response has been acquired independently under contact and non-contact conditions (Fig. 4b inset), showing a frequency shift that corresponds to that measured during the approach-retract cycle. While in contact, the mechanical Q-factor experiences a 4-fold decrease compared to non-contact, which indicates dissipative contact interactions.

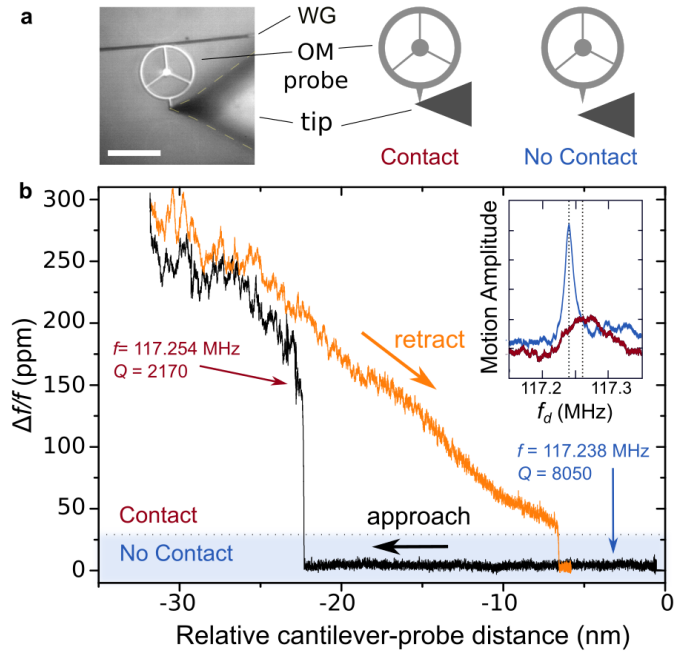

**Figure 4 | Approach-retract experiment. a,** Optical photography of the commercial cantilever tip approaching the optomechanical (OM) probe (left panel). Scale bar: 20 μm. Schematic of the experiment where the cantilever tip is in contact (middle panel) and non-contact (right panel), WG: optical waveguide. **b,** Normalized mechanical frequency shift versus relative tip-probe distance in an approach-retract cycle. Inset: Difference of frequency and mechanical Q-factor in contact (red) or not (blue).

To finally demonstrate the large bandwidth of the optomechanical force probe, a time-modulated electrostatic force, inducing a force gradient, is generated between the probe apex and the cantilever tip by applying a modulated voltage at frequency $f_{Force}$ between them. The tip is positioned a few nanometres before contact, the modulation frequency ranges from 2 kHz to 1 MHz, and we measure the probe vibration amplitude in response to this stimulation. The frequency response of the probe to this modulated force gradient is shown in Fig. 5. The response is flat up to a cut-off frequency of 26 kHz, of order $\Gamma$ the invert of the amplitude ring down time corresponding to the measured mechanical $Q$~2790 ($\Gamma = 2\pi f/Q$). With the here employed protocol, the bandwidth of the force measurement is close to $f/Q$, but it can in principle approach $f$ [12]. While $Q$ can be controlled by tuning the size of the pedestal, by implementing a feedback control, or by operating in a liquid[23], it is really the probe frequency that sets the perspective in force bandwidth. Our results indicate that it could be pushed to the very high frequency range with the here-demonstrated approach.

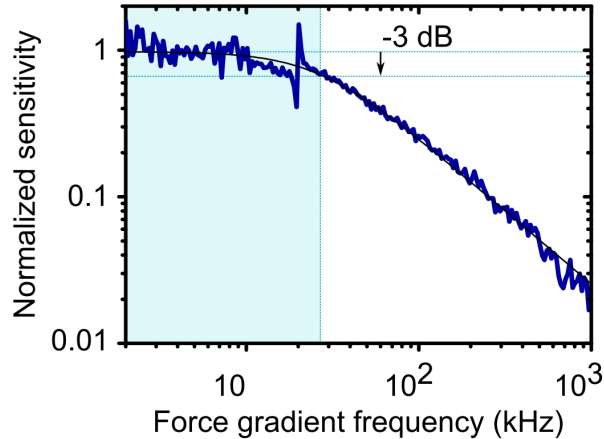

**Figure 5 | Normalized frequency response of the optomechanical probe amplitude to a modulated electrostatic force gradient.** The measurements show a low-pass filter behaviour with a cut-off frequency of 26 kHz.

## Discussion

In summary, we have developed a new mechanical atomic force probe that operates in sub-picometre oscillating mode at frequencies above 100 MHz, bringing orders of improvement over state-of-the-art AFM probes. On top of these assets, and in contrast to flexural nanobeam probes evanescently coupled to an optical cavity[24,25], the strong optomechanical coupling enables here all-optical operation of the probe, both in vibration actuation and detection. This permits low-noise force acquisition in the important mechanically resonating mode, with a large dynamic range merely limited by thermomechanical noise.

With this resonating probe and its nanoscale tip-shaped apex, we have demonstrated sensing of contact forces, as well as very high speed sensing of electrostatic interactions. Our results not only prove the validity of the approach but also indicate that the force bandwidth could be pushed up to the GHz with optimized versions of the probe[17,18], a range that has remained out of reach of mechanical sensing. An obvious application is higher-speed AFM imaging, as the here demonstrated concept already unlocks limitations associated to current beam cantilevers. At the single pixel level though, great many investigations would already become accessible with such force bandwidth. This spans condensed matter, with the dynamics of domains in glass-forming bodies and of vortices in superconductors; and biology, with conformational changes in cell membranes and molecules, where the ultra-fast spectroscopy of dissipation may elucidate relations between conformation and function.

## Acknowledgements

The authors acknowledge support from the French Agence Nationale de la Recherche through the Olympia ANR-14-CE26-0019 project and from the French DGA. P. E. Allain and I. F were supported by the European Research Council through the GANOMS project n°306664. The authors thank Eduardo Gil-Santos and Denis Lagrange for discussions.

## Author contributions

S. H, G. J, B. L and I. F conceived the principles of the probe. P. E. A, S. H, M. F, G. J, B. L and I.F established the appropriate design. C. M, M. G, E. M, M. H, B. W and G. J. led the fabrication. P. E. A, L. S, G. L, B. L and I. F planned experiments. P. E. A and L. S carried-out systematic experiments. P. E. A, L. S, B. L and I. F interpreted experiments, while all authors discussed implications of the results and contributed to the manuscript.

## Methods

**Device Fabrication.** The optomechanical resonators and waveguides were patterned from a SOI wafer with e-beam lithography followed by an inductively coupled plasma reactive ion etching (ICP-RIE) step. Electron dosage was fine-tuned to compensate proximity effects. For some devices, the die was cleaved at a distance of 1 mm from the resonator to terminate the waveguides and provide optical coupling conditions at the cleaved facets. Finally, the resonators were released by HF vapour etching of the buried silicon oxide layer. Images of devices fabricated using the VLSI technology can be found in Fig. S4 of the Supplementary Information.

**Design**. For simplicity, the apex has been diametrically opposed to the spokes but it can be positioned differently if willing to increase its displacement while reducing its stiffness (see Supplement Information). In Fig 2a's design, the simulated spring constant of the resonator taken at the apex reduction point is $k \sim 2.6 \times 10^6$ N/m. If the spokes are rotated 30° anticlockwise, the spring constant is $k \sim 4$ kN/m while the high optical quality factor is preserved.

**Optical driving of the probe.** Electro-Optical Modulators were operated to have a sinusoidal response to the applied voltage. A maximal driving voltage of 750 mV was chosen to remain in the linear regime and hinder harmonics generation. The transmission of such modulator being sensible to temperature, the latter was regulated. In order to actuate effectively the mechanical motion, the optical wavelength was set practically at resonance to the optical mode.

**Measurement of the frequency shift deviation.** See Fig. S3 of the Supplementary Information for definitions. To estimate the frequency shift $\Delta f$ without locking, the phase $\theta_1$ between the drive signal and the resonator response, extracted from the signal vector exiting the LIA, was converted to the frequency knowing in advance the slope of the phase rotation versus drive frequency: $\alpha = 4.42 \cdot 10^{-4}$ °/Hz. The measured frequency shift was thus $\Delta f/f = (\theta_1 - \theta_{1\_initial}) \times \alpha/f_0$. The driving voltage was set to 200 mV and the LIA filter BW was set to 100 kHz. As expected from white noise limited experiment, the deviation scaled as $\tau^{-1/2}$ in the region $\tau < 10$ s.

**Force Curve experiment.** Gold was sputtered on the cantilever tip and the gold layer was wire-bonded to remotely set its electrical potential. In phase-locked loop configuration, the measured frequency shift is $\Delta f/f = (f - f_0)/f_0$ with $f = f_d$.

**Frequency response of the optomechanical probe to an external force gradient.** A modulated (from 2 kHz to 1 MHz) electric potential was applied to the golden tip. For the whole experiment, the probe was driven at its maximum amplitude ($V_d = 750$ mV @ 117 MHz). To ensure that the response of the probe amplitude was purely electrostatic and not interfered by other electrical or opto-electrical phenomena, the demodulated signal $R_2 e^{i\theta_2}$ (see Supplementary Information and Fig. S3) was acquired in two configurations: I) $R_{2\_IN} e^{i\theta_{2\_IN}}$: on the flank of the mechanical mode amplitude of maximum slope with $f_d = f_0 - (\Gamma/2\pi)/2$; II) $R_{2\_OUT} e^{i\theta_{2\_OUT}}$: far from the mechanical resonance with $f_d = f_0 - 65 \times \Gamma$. Finally the normalized transfer function $H_{\_LPF}$ of the demodulation low pass filter of Lock-in 1 (BW= 500 kHz) was acquired. The low pass filter compensated probe amplitude was then $|(R_{2\_IN} e^{i\theta_{2\_IN}} - R_{2\_OUT} e^{i\theta_{2\_OUT}})/H_{\_LPF}|$. We checked that the measurement chain bandwidth (cables, photodetector) was much higher than 1 MHz. The "Fano-like" peak at 20 kHz is interpreted as a mechanical mode of the cantilever.

**Data availability.** The data that support the findings of this study are available from the corresponding author upon reasonable request.

# Supplementary Information

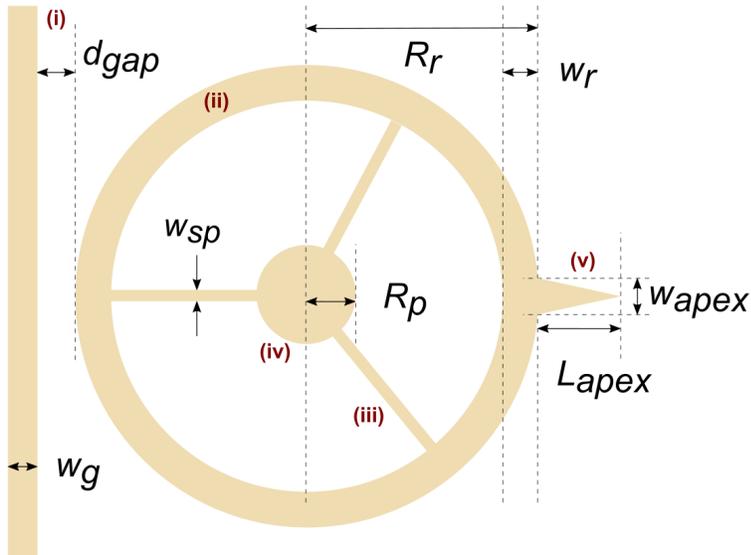

**Sup. 1 | Details of the probe design**. (i) Waveguide. (ii) Ring optomechanical resonator. (iii) Spokes. (iv) Pedestal. (v) Probe apex. $w_g$: width waveguide. $d_{gap}$: gap between waveguide and ring. $R_r$: Ring radius. $w_r$: ring width. $w_{sp}$: width of the spokes. $R_p$: pedestal radius. $w_{apex}$: base width of the OM probe apex. $L_{apex}$: Length of the OM apex. For experiments presented in the main text: $w_g$= 400 nm, $d_{gap}$= 300 nm, $R_r$= 10 μm, $w_r$= 750 nm, $R_p$= 1.5 μm, $w_{sp}$=$w_{apex}$= 500 nm, $L_{apex}$= 4 μm.

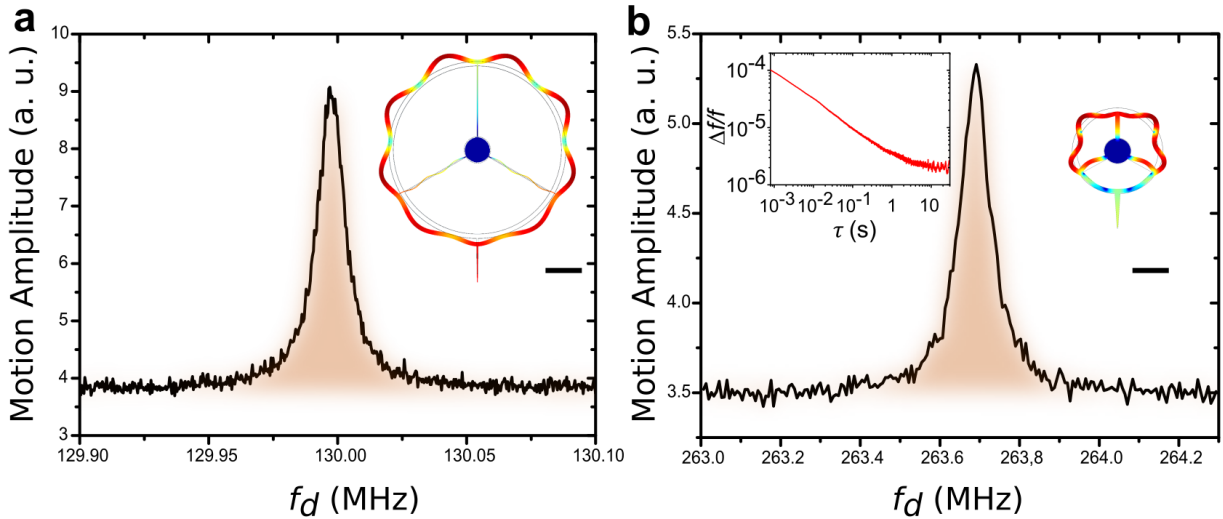

**Sup. 2 | Motion amplitude ($R_1$) of other probes characterized in the Brownian regime ($V_d$=0 V). a**, Softer probe $f$=130 MHz, $k\sim$7 kN/m, $R_r$=10 μm, $w_r$=500 nm, $w_{sp}$=100 nm, $w_{apex}$=100 nm, $L_{apex}$= 4 μm, Filter BW=1kHz. **b**, Higher frequency probe $f$=263.7 MHz, $k\sim$143 kN/m. $R_r$=5 μm, $w_r$=500 nm, $w_{sp}$=500 nm, $w_{apex}$=100 nm, $L_{apex}$= 4 μm, Filter BW=1 kHz. <u>Inset:</u> Allan deviation of the normalized mechanical frequency shifts $\Delta f/f$ showing the relative accuracy of the probe's measurement in open loop for $V_d$=200 mV. BW=100 kHz. Scale bars: 3 μm

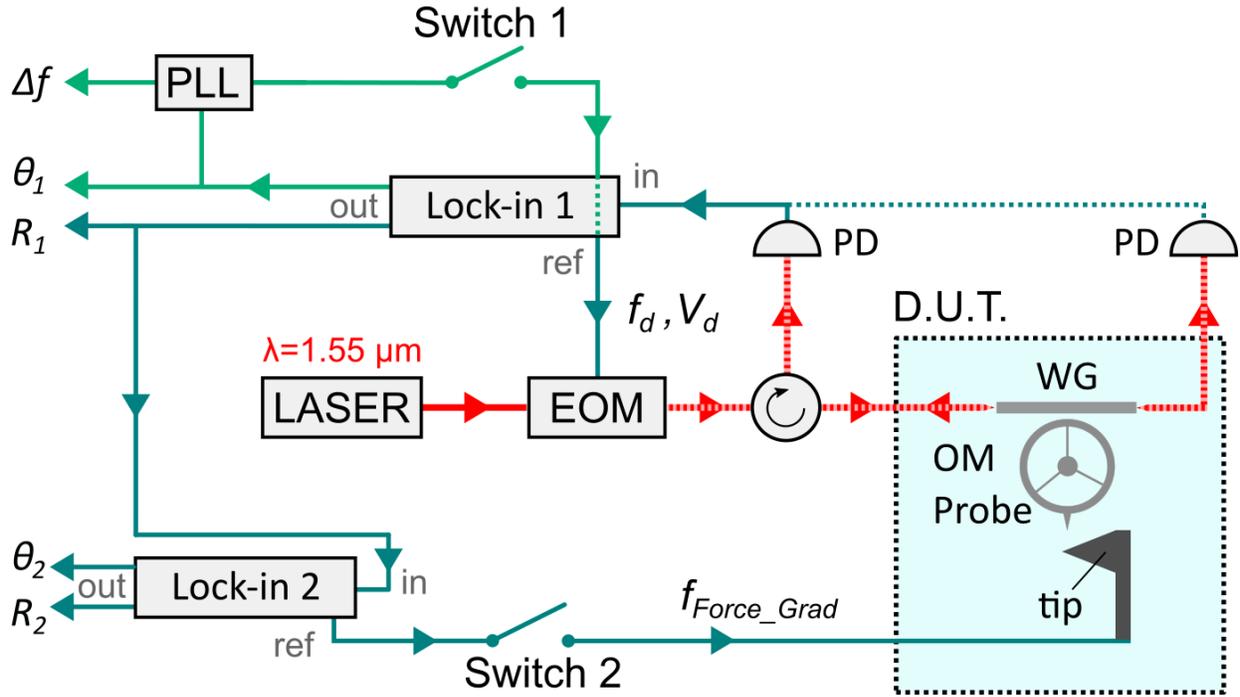

**Sup. 3 | Bloc diagram of the experimental setup.** Laser light used to excite the optomechanical (OM) probe is modulated by an electro-optic modulator (EOM) driven at frequency $f_d$. The device under test (DUT) can be accessed either in optical reflection or transmission. The experiments reported in the main text were carried in reflection from the waveguide, using a fibered optical circulator. The information imprinted by the OM probe motion onto the collected light is converted into a RF signal by a photodetector (PD) and sent to a lock-in amplifier (Lock-in 1). The complex component $R_1 e^{i\theta_1}$ at $f_d$ of this signal exits Lock-in 1. In phase-locked loop (PLL) configuration (switch 1 closed), $\theta_1$ is fed-back to a PLL instrument to lock $f$ to $f_d$. In absence of phase-locking (Fig. 3c), $\theta_1$ is converted to $\Delta f$ using the slope of the phase rotation versus frequency. In the force modulation experiment, $R_1$ is fed to Lock-in 2. The component of $R_1$ at the force modulation frequency $f_{Force}$ is extracted: in Fig. 5 $R_2$ is plotted versus $f_{Force}$.

| Figure | Plotted quantity | | | Switch 1 | Switch 2 | $V_d$ | |
|---|---|---|---|---|---|---|---|
| Fig. 3 | **a:** R1 | **b:** max(R1) | **c:** $(f-f_0)/f_0$ | open | open | **a, b:** 0→750 mV | **c:** 200 mV |
| Fig. 4 | $(f-f_0)/f_0$ with $f=f_d$ | | | closed | open | 200 mV | |
| Fig. 5 | $\lvert (R_{2\_IN} e^{i\theta_{2\_IN}} - R_{2\_OUT} e^{i\theta_{2\_OUT}})/H_{\_LPF} \rvert$ | | | open | closed | 750 mV | |

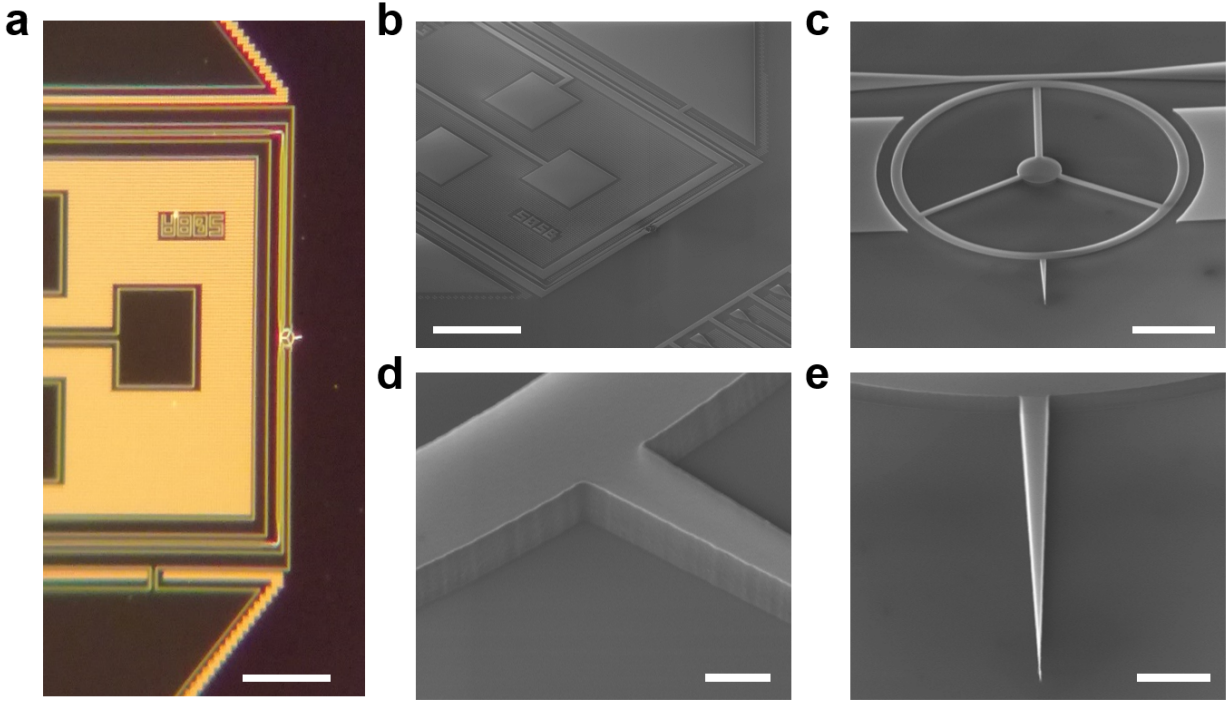

**Sup. 4 | Images of devices fabricated using the VLSI technology**. **a**, Optical microscope top view of a boat-shaped ultra-fast optomechanical probe device. Scale bar: 116 µm **b**, SEM monograph of the device bow. Scale bar: 100 µm **c**, Zoomed image of an optomechanical ring probe. Scale bar: 3 µm **d**, Zoomed image of the apex's base before under-etching. Scale bar: 400 nm **e**, Zoomed image of the apex. Scale bar: 800 nm. $w_g$= 620 nm, $d_{gap}$= 100 nm, $R_r$= 10 µm, $w_r$= 750 nm, $R_p$= 1.5 µm, $w_{sp}$= 400 nm, $w_{apex}$= 400 nm, $L_{apex}$= 5 µm.